\documentclass[aps,prl,twocolumn,groupedaddress,showpacs]{revtex4}
\usepackage{graphicx}       % Include figure files
\usepackage{dcolumn}        % Align table columns on decimal point
\usepackage{bm}             % bold math

\newcommand{\rmsd}{$\langle \Delta \bar{u}_{\bot}^2 \rangle$ }	% <u-bar-squared>

\begin{document}

% ---------For Line numbering in the document----------------------------------
% \linenumbers							
% \def\linenumberfont{\normalfont\small\sffamily}
%------------------------------------------------------------------------------

%------Title & Author Lists----------------------------------------------------
\title
    {
        Electronically driven fragmentation of Ag nanocrystals
        revealed by ultrafast electron crystallography
    }

\author{Ramani K. Raman}
\affiliation{Physics and Astronomy Department,
             Michigan State University,
             East Lansing, Michigan 48824-2320 }

\author{Ryan A. Murdick}
\affiliation{Physics and Astronomy Department,
             Michigan State University,
             East Lansing, Michigan 48824-2320 }

\author{Richard J. Worhatch}
\affiliation{Physics and Astronomy Department,
             Michigan State University,
             East Lansing, Michigan 48824-2320 }

\author{Yoshie Murooka}
\affiliation{Physics and Astronomy Department,
             Michigan State University,
             East Lansing, Michigan 48824-2320 }

\author{Subhendra D. Mahanti}
\affiliation{Physics and Astronomy Department,
             Michigan State University,
             East Lansing, Michigan 48824-2320 }

\author{Tzong-Ru T. Han}
\affiliation{Physics and Astronomy Department,
             Michigan State University,
             East Lansing, Michigan 48824-2320 }

\author{Chong-Yu Ruan}
\email[]{Email: ruan@pa.msu.edu}
\affiliation{Physics and Astronomy Department,
             Michigan State University,
             East Lansing, Michigan 48824-2320 }

%\date {	\today}

%------------------------------------------------------------------------------

%-------------Abstract---------------------------------------------------------
\begin{abstract}

We report a ultrafast electron diffraction study of silver nanocrystals under
surface plasmon resonance excitation, leading to a concerted fragmentation.
By examining simultaneously transient structural, thermal, and Coulombic
signatures of the prefragmented state, an electronically driven nonthermal
fragmentation scenario is proposed.

\end{abstract}
%------------------------------------------------------------------------------

%---------PACS Numbers, listed as per 2008 scheme------------------------------
\pacs
{		
			36.40.Gk,		% Plasma & collective effects in clusters
			73.22.Lp,   % Collective Excitations - Electronic structure of
									% nanoscale materials: clusters, nanoparticles, nanotubes,
									% and nanocrystals
			78.67.Bf, 	% Nanocrystals & nanoparticles - Optical properties of
									%	low-dimensional, mesoscopic, and nanoscale materials and
									%	structures
			61.05.J- 		% Electron diffraction and scattering  	
		% 79.60.-i,		% Photoemission
  	%	81.05.Uw, 	% Materials science -- Carbon, Diamond, Graphite
  	%	52.59.Sa,		% Space-charge-dominated beams (plasmas),
  	%	61.46.Hk,		% Structure of nanoscale materials - 	Nanocrystals
  	%	61.72.Ff,		% Direct observation of dislocations and other defects
  	%							%	(etch pits, decoration, electron microscopy,
  								%	x-ray topography, etc.)
		% 65.40.De  	% Thermal expansion; thermomechanical effects
		% 82.53.Mj 		% Femtosecond probing of nanostructures
		% 73.22.-f 		% Electronic structure of nanoscale materials: clusters,
		%           	%	nanoparticles, nanotubes, and nanocrystals
		% 78.20.Bh 		% Optical properties: Theory, models, and numerical
									%	simulation
		% 71.15.Qe 		% Excited states: methodology
		% 68.05.Cf 		% Surface Structure: measurements and simulations
		% 68.35.Ja 		% Surface and interface dynamics and vibrations
		% 71.20.-b 		% Electron density of states and band structure of
			 						%	crystalline solids
		% 78.66.Tr 		% Optical properties: Fullerenes and related materials
		%	82.30.Lp 		% Decomposition reactions (pyrolysis, dissociation, and
									%	fragmentation)
}

% \keywords{}
%------------------------------------------------------------------------------

\maketitle

%------------------------------------------------------------------------------

%----Environment for displaying long equations, when in 2-column mode---------
% If in two-column mode, this environment will change to single-column
% format so that long equations can be displayed. Use sparingly.
%\begin{widetext}
%\end{widetext}
%------------------------------------------------------------------------------

The ability to image defect growth processes is central to the understanding
of the electronically induced structural phase transitions in solids
\cite{Nasu04, ColletScience03, QazilbashScience07}. Whereas optical and
photoemission studies have provided significant insights into the initial
electronic processes that are strongly coupled to lattice degrees of freedom,
the mechanism bridging the femtosecond (fs) optical seeding to the
picosecond(ps)-to-nanosecond(ns) macroscopic structural changes remains a
central topic to be elucidated. Recent developments in ultrafast diffraction
techniques have enabled direct probing of atomic dynamics and helped accentuate
the important role of electronic excitation in initiating coherent motions
\cite{BargheerScience04}, bond dilation \cite{BaumScience07, FritzScience07} and
structural transformations \cite{RuanScience04,GedikScience07,SiwickScience03}.
Here, employing ultrafast electron nanoscale crystallography \cite{RuanMM09},
we demonstrate a direct structural study of spatially inhomogeneous processes
in Ag nanocrystals (NCs) induced via surface plasmon resonance (SPR) excitation.
Contrary to an impulsive process leading to fragmentation, we find that the
dominant dynamical feature in the prefragmentation stage is a defect-mediated
instability growth, creating sub-nanocrystalline domains with hot surface and
relatively cold core.  Electronic effects are proposed to account for the
incipient creation and subsequent growth of lattice inhomogeneities directly
responsible for fragmentation, which are corroborated by the evidences of
correlated charge localization and defect percolation on the picosecond
timescale following photoexcitation.
%------------------------------------------------------------------------------
%\newpage
% entire page. There is no need to do explicit centering.
\begin{figure}
	\includegraphics[width=1.0\columnwidth]{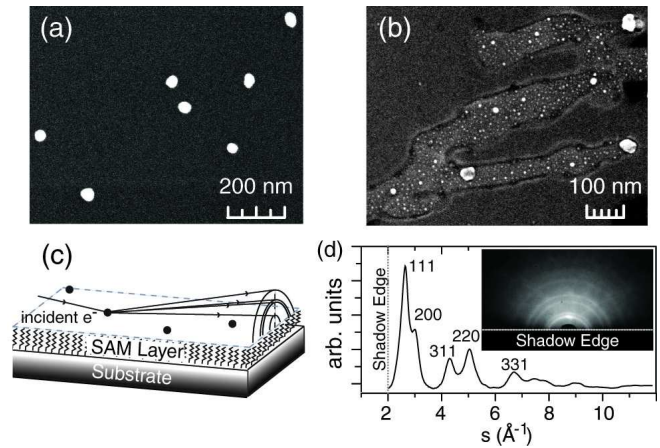}
	\caption{	(a) SEM images of 40 nm Ag NC sample
								before laser irradiation and
						(b) after irradiation (400 nm, $F=$ 22 mJ/cm$^2$), showing
								fragmentation.
						(c) Schematics of diffraction/sample geometry employed in this study.
            (d) Ground state 1D structure function obtained from the 2D
            		powder diffraction pattern (inset), showing characteristic fcc
            		peaks.
				}\label{Fig1}
\end{figure}

%------------------------------------------------------------------------------
Photoinduced fragmentation of NCs has been a subject of recent intense
interests, with a range of results favoring either thermal ablation
\cite{TakamiJPCB99,InasawaJPC05}, or a nonthermal pathway involving charging
of the nanoparticles \cite{KamatJPCB98, LinkJPCB00, RedmondJPCC07a, MutoJPCC08}
or creation of strong fields \cite{PlechNPhys06}. The mechanism for photoinduced
structural changes appears to depend on the pulse duration \cite{MazurMST01},
with the surrounding solvent limiting the thermal dissipation
\cite{HartlandPCCP04}. Using ultrafast electron crystallography (UEC)
\cite{RuanMM09}, we have directly probed the transient structures of the NCs up
to the ablation limit, and identified the transient dynamical structures and the
charge states. In contrast to earlier colloidal suspensions studies, this
experiment is conducted in vacuum with sufficient dispersion of
nanoparticles to avoid long-range surface plasmon induced interaction and
solvent effect. Colloidal silver NCs (\emph{Ted Pella}, 40$\pm$7 nm) are dispersed
\emph{ex-situ} on a silicon substrate functionalized with a self-assembled
monolayer (SAM) of alkane-silane ((3-Aminopropyl)trimethoxysilane or APTMS,
\emph{Ted Pella}) molecules that suppress substrate signals. SEM
characterization of the sample showed uniform coverage, with low mean areal
density of 7 NC per $\mu m^2$ (Fig.~\ref{Fig1}(a)). First, the NCs are irradiated
using ultrafast laser pulse (50 fs, \emph{p}-polarized, 45$^{\circ}$ incidence)
at 400 nm, which is within their SPR bandwidth ($385\pm40$~nm), to check for
traces of fragmentation. At a critical fluence F=22 mJ/cm$^2$ (multi-shot) SPR excitation
leads to fragmentation into predominantly 2 nm NCs, as shown in
Fig.~\ref{Fig1}(b). Prior to such laser excitation, the UEC patterns obtained
from these randomly oriented NC samples resemble a `powder diffraction' pattern,
as shown in Fig.~\ref{Fig1}(c). The 1D structure function deduced via a radial
averaging of these 2D patterns \cite{RuanMM09} displays clear fcc symmetry
(Fig.~\ref{Fig1}(d). We subsequently probe the transient structures and the
atomic rearrangements using laser fluence \emph{just below} the critical
fragmentation fluence. Sample integrity was retained during pump-probe cycles
of UEC, as confirmed by SEM imaging following the conclusion of experiment.
%------------------------------------------------------------------------------
%\newpage
% entire page. There is no need to do explicit centering.
\begin{figure}
	\includegraphics[width=1.0\columnwidth]{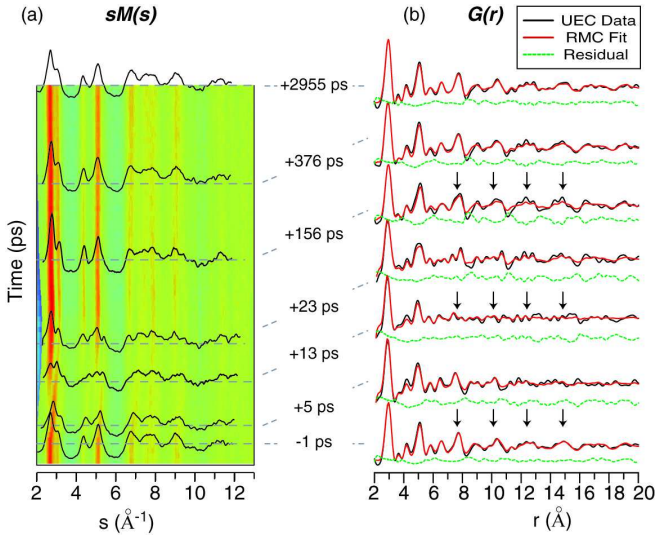}
	\caption{ (Color online) UEC Data.
						(a) Time evolution map of the normalized structure factor
								\emph{sM(s)} of Ag NCs excited at SPR at $F=$ 17 mJ/cm$^2$,
								with \emph{sM(s)} at select times explicitly
								overlayed.
						(b) Pair correlation function $G(r,t)$, corresponding to the
								select times in (a). The arrows point to the medium range
								order peaks, which completely dephase by $t=+$ 13 ps, before
								their eventual revival at long times.
				}\label{Fig2}
\end{figure}
%------------------------------------------------------------------------------

The gross features of photoinduced structural changes are evident from the
transient alteration of the normalized structure function \cite{RuanMM09}
$sM(s)$, ($s$ is the electron scattering wave-vector) obtained
at the prefragmenting fluence $F=$ 17 mJ/cm$^2$, shown in Fig.~\ref{Fig2}(a).
By $t=+$13 ps, the initial fcc ordered diffraction maxima diminish while the
nearby diffusive scattering gain strength and grow into multiple peaks,
indicating a departure from the cubic symmetry. The reduction of long-range
order is evident from the corresponding pair correlation function
$G(r,t)$ \cite{RuanMM09}, obtained via a Fourier analysis of the diffraction
patterns, as shown in Fig.~\ref{Fig2}(b). The atom-atom correlation peaks
greater than 10 {\AA} in $G(r,t)$ smear out completely by $t=+$13 ps, whereas
the short-range peaks ($r \leq$ 10 {\AA}) retain their strength, thereby
indicating the persistence of short-range order during the
pre-fragmentation process.
%------------------------------------------------------------------------------
%\newpage
% entire page. There is no need to do explicit centering.
\begin{figure}
	\includegraphics[width=1.0\columnwidth]{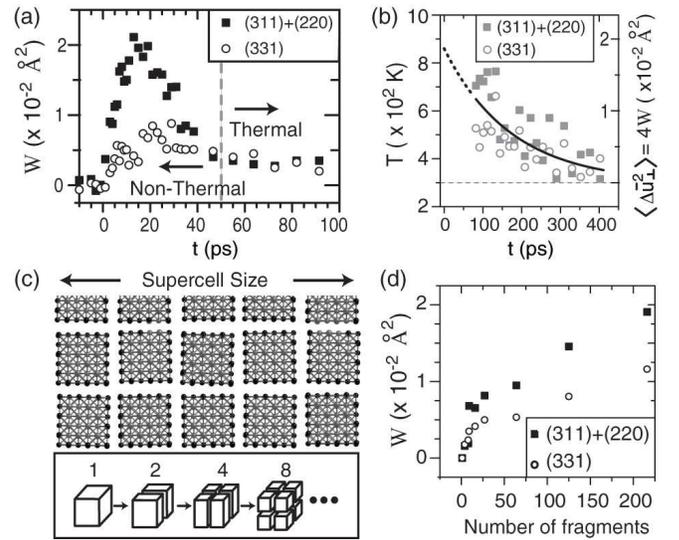}
	\caption{Thermal signatures of fragmentation process.
						(a) $W = (1/s^2)$ ln $(I/I_0)$ determined from intensity drop of
								diffraction rings and
						(b) corresponding lattice temperature deduced at long times,
								with exponential fit (solid line) to data beyond $t>80$~ps.
								Early time data are excluded for clarity. See \cite{EPAPS}.
						(c) Planar slice of the fragmented supercell. Random relative
								orientations imposed amongst the fragments removes inter fragment
								correlation. Black atoms indicate undercoordinated sites. The
								cartoon depicts the fragmentation scheme.
						(d) Variation of simulated $W$ with the number of fragments.
					}	\label{Fig3}
\end{figure}
%------------------------------------------------------------------------------

To identify the origins of this structural disorder, we first calculate from
each Bragg intensity a parameter $W=(1/s^2)$ ln $(I/I_0)$. $W$ is closely
related to the mean-square displacement of atoms \rmsd (4W=\rmsd) - a tool
often used in cases of quasi-equilibrium conditions to extract lattice
temperature via the Debye-Waller relation. In such cases, \rmsd is independent
of the diffraction order used \cite{RamanPRL08}. However, we find here
(Fig.~\ref{Fig3}(a)) that for $t<50$ ps, $W$ determined from higher order
(331) peak is smaller than that from the lower order (311) and (220)
peaks, indicating the nonthermal nature of the structure disordering process.
We will argue that this anisotropy in $W$ is the manifestation of a unique
structural change that destroys the long-range coherence within NCs acting
as a precursor to the fragmentation observed above the critical fluence. At
longer times ($t>50$~ps), $W$ shows no $s$-dependence, thus allowing us to
reliably extract the lattice temperature from $W$ \cite{RamanPRL08}, shown
in Fig.~\ref{Fig3}(b). The temperature decay is in good agreement with an
interface limited thermal relaxation model based on thermal conductivity
of the SAM \cite{EPAPS} and suggests a temperature rise in NCs of at most
$860\pm150$ K at early times, which is insufficient to cause thermal ablation.
To capture the essence of this nonthermal transformation, we carry out a
simulation of diffraction signatures expected from NC fragmentation.
Starting from a $15\times15\times15$ supercell structure for the NC,
we fragment (subdivide) the supercell into smaller units as illustrated
in Fig.~\ref{Fig3}(c). The separation between fragments and their relative
orientations is altered randomly so as to remove any inter fragment
correlations. Such fragmented assemblies are generated for several
fragmentation numbers, following which, the $sM(s)$ intensity from the
fragmented assembly is computed for each case. The lattice temperature is
maintained (at 300 K), so that any changes evident in the $sM(s)$ intensities
arise solely from the fragmentation process. The $W$ parameter extracted from
these `simulated' $sM(s)$ is shown in Fig.~\ref{Fig3}(d), and exhibits the same
anisotropy trend seen in the experiment. A qualitative agreement in the scale
of anisotropy seen in the experiment and simulation is reached when the number
of fragments exceed 100, corresponding to an individual fragment size of
$\approx 2$~nm, which is close to the average fragment sizes observed in the
SEM image (Fig.~\ref{Fig1}(b)). These results indicate that a nonthermal
restructuring (prefragmentation) rather than a thermal disorder better
explains the experimental observations.

\begin{figure}
	\includegraphics[width=1.0\columnwidth]{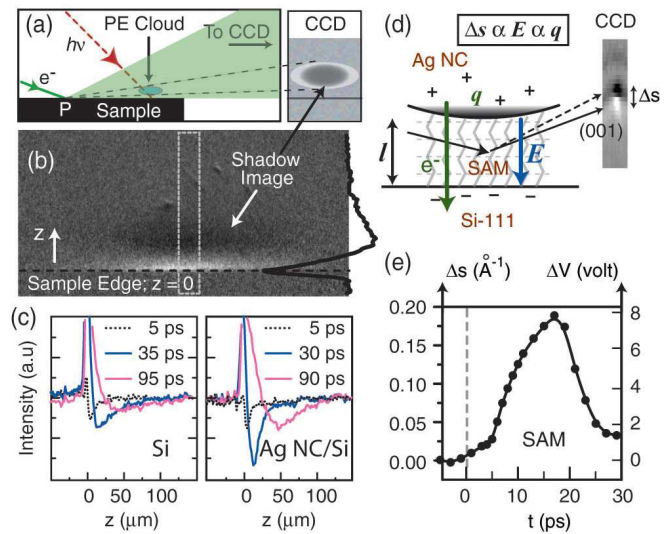}
	\caption{ (Color Online) Charge signatures of fragmentation process.
			(a) Point-projection imaging of vacuum emitted photoelectrons (PE).
			(b) Normalized shadow images of the PE cloud. The cloud profile is shown
					on the right.
			(c) Evolution of PE cloud profiles over Si-111 surface with and
					without the Ag NCs, showing significant enhancement in presence of
					Ag NCs.
			(d) Concept of diffractive voltammetry.
			(e) Shift of the SAM (001) peak with time and corresponding interfacial
					potential $\Delta V = E \cdot l$, using $l=1$~nm.
			}	\label{Fig4}
\end{figure}

To find out the role of Coulombic forces in the fragmentation of the NCs, we
examine two relevant processes leading to charging of the NCs: photoemission
and charge transfer between the NCs and the supporting surface. To determine
the photoemission yield, we use a projection imaging method \cite{RamanAPL09}
illustrated in Fig.~\ref{Fig4}(a,b), where surface scattered electrons originating
from point P intercept the laser-induced vacuum emitted charge cloud, thus
casting its shadow on the CCD. By analyzing the normalized shadow images of
the electron cloud, we can extract the density of photoemitted electrons, which we
find to double (to 7.9$\times 10^7$ e/cm$^2$) in the presence of NCs
relative to that from a bare Si surface, as evident from Fig.~\ref{Fig4}(c).
This enhancement is significant considering the NCs occupy less than $1\%$
of the surface. Similar photoemission enhancement from Ag NCs has been
observed in other studies as well and attributed to SPR assisted multiphoton
processes \cite{LehmannPRL00}. In addition to this, we observe another competing,
SPR enhanced localized charge transfer channel between NCs and the substrate,
that causes even more pronounced charging of the NCs. We quantify this process
via a diffractive voltammetry approach \cite{RuanMM09,MurdickPRB08, RamanAPL09}
as illustrated in Fig.~\ref{Fig4}(d). Recall that the Ag NCs sit on top of a SAM layer,
which produces its own diffraction signal in the form of a strong (001) peak at
$s=$2.75 {\AA}$^{-1}$. Any interfacial charge transfer between the Ag NC and
the substrate leads to the creation of a transient field at the Ag NC/SAM/Si
interface, causing a refraction shift of probing electron beam traversing the
region in between. Thus, from the shift of the (001) SAM peak, we numerically
compute the corresponding interfacial field $E$ (and potential $\Delta V$).
Based on an interfacial capacitance C=7.5 attofarad, we thus determine a maximum
average charging per NC of $q \approx$ 400 e$^+$ , which is in contrast to 0.2 e$^+$
estimated from the vacuum emission measurement. This corresponds to a Coulomb fissility
ratio $X=q^2/n \approx 0.08$, where $n \approx 2\times 10^6$ is the number of atoms
within the NC, which is below the direct Coulomb fissility regime
 ($0.3<X<1$) \cite{MutoJPCC08}.

%------------------------------------------------------------------------------
%\newpage
% entire page. There is no need to do explicit centering.
%------------------------------------------------------------------------------

Our observation is consistent with a progressive Coulomb-induced fragmentation
scenario proposed by Kamat and coworkers \cite{KamatJPCB98}. The transient
charging curve (Fig.~\ref{Fig4}(e)) shows an initial 6 ps incubation period for
charge redistribution at NC interface which is absent at non-SPR excitations
\cite{RuanMM09}, and is thus indicative of transient trapping of positive
charges at the excitation sites. The migration of charges to the interface
and the accompanied rise of interfacial field (Fig. \ref{Fig4}(e)) is hindered
as the lattice is undergoing atomic restructuring during this incubation period,
which also coincides with the period of $W$ - anisotropy (Fig \ref{Fig3}(a)).

The observed correlation between the atomic process and charge trapping
suggests that local valence instabilities which cause bond softening
\cite{FritzScience07} lead to structural defects and charge localization as
seen in our studies. Earlier studies of fs laser induced melting in metals using
optical and photoemission techniques have revealed that interband transitions
\cite{GuoPRL00} and thermionic emission \cite{GruaPRB03} can both lead to
rapid bond softening within the initial non-equilibrium time scales of core hole
lifetime ($\approx$ 100 fs) and electron-phonon coupling time ($\approx$ 1 ps)
\cite{DelFattiPRB00}.  In Ag, the nonlinear interband excitation involving
inner valence shell ($d \rightarrow sp$) is found to be strongly coupled to
the SPR dephasing pathway \cite{VoisinJPCB01}. These excitations can provide
seeds for valence destabilization through SPR. Such a process would be
inhomogeneous in nature and proceed on a phonon timescale.

%------------------------------------------------------------------------------
%\newpage
% entire page. There is no need to do explicit centering.
\begin{figure}
	\includegraphics[width=1.0\columnwidth]{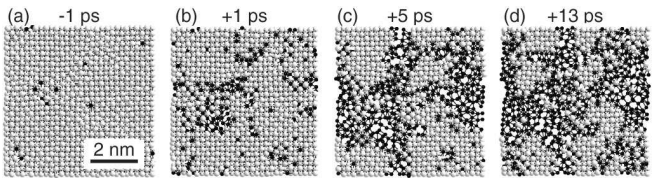}
	\caption{ Prefragmentation states of Ag NC at select times,
						determined using the structure refinement scheme described in the text.
						Growth of local disorder is apparent from the increase in number
						of undercoordinated atoms (black)
	 }	\label{Fig5}
\end{figure}
%------------------------------------------------------------------------------
To visualize transient atomic processes associated with electronic excitation,
we employ a Reverse Monte Carlo (RMC) sturcture refinement scheme
\cite{RuanMM09}. While the earlier fragmentation simulation yielded qualitative
insights into the origin of observed $W$ - anisotropy, the approach here seeks to
refine/adjust the local atomic positions within the $15\times15\times15$
Ag supercell so as to fit both the positions and intensities of $G(r,t)$
peaks generated from these structures to those determined experimentally
at each time instance, as shown in Fig.\ref{Fig2}(b). First, we confirm
that the data has sufficient signal-to-noise ratio since the RMC result of
the ground state ($t<$0) retains a robust fcc sturcture, as shown in
Fig.~\ref{Fig5}(a). Given that the diffraction changes between consecutive
time frames are small, the refinement scheme progressively tracks the
disordering process from one time instance to the next, while being constrained
by the correlation between atomic positions in neighboring time frames that is
inherently present in the diffraction patterns. Selected RMC results are
depicted in Figs.~\ref{Fig5}(a)-(d) with a planar slice cut off from the 3D
supercell
to expose the degradation of the pristine fcc lattice over time. The atomic
sites colored in black represent topological defects with coordination number
N$_c \leq$ 10, whereas the fully coordinated ones have N$_c=$ 12 corresponding
to the fcc structure. We find that these defects, which are initially sparsely
distributed following SPR, visibly grow and percolate into strips in 10 ps and
then saturate, leaving behind fragmented nanocrystalline domains with an average
size $\approx 2$~nm \cite{EPAPS}. This disordered state persists for nearly 10 ps
before a slow recovery to fcc on a time scale $\approx 100$~ps.  Further analysis
shows that among those undercoordinated sites the nearest neighbour distance
exhibits large fluctuations indicating bond-softening in this region
\cite{FritzScience07} characterized by their averaged root-mean-square
displacements on the order of 0.15 {\AA}, which is 4 times higher than those
in the region of core domains.

In summary, an electronically driven, progressive fragmentation of Ag NCs is presented.
We suggest that the fragmentation process is triggered by the creation of local valence
instabilities, caused by redistribution of local charge density, facilitated in Ag
through the strong nonlinear coupling between SPR and interband transition. When
sufficient valence instabilities are instigated at a high fluence
($F \geq$ 17 mJ/cm$^2$), the electronic states are strongly perturbed, leading to a
prolonged ps lifetime of the local charge redistribution associated with valence
excitation due to an insufficient dynamical screening. Such a dynamical localization
feature is central to the creation and growth of the topological defects, which can
persist on the phonon timescales, and could be common in nonequilibrium photoinduced
structural phase transition.

We thank P.M. Duxbury, S.J.L. Billinge, R. Shen, and F. Vall\'{e}e for critical
discussions.  The experimental work was supported by Department of Energy under
Grant DE-FG02-06ER46309. The theoretical analysis was supported (R.J.W, R.A.M) by
National Science Foundation under Grant DMR-0703940.
%+++++++++++++++++++++++++++++++++++++++++++++++++++++++++++++++++++++
% \bibliographystyle{apsrev}% your bst file here
% \bibliography{AgNP09}%your bib file here

\end{document}